# Compressive failure as a critical transition:
# Experimental evidences and mapping onto the universality class of depinning

Chi-Cong Vu[1], David Amitrano[1], Olivier Plé[2] and Jérôme Weiss[1,*]

[1]Univ. Grenoble Alpes, CNRS, ISTerre, 38000 Grenoble, France
[2]Univ. Savoie Mont-Blanc, CNRS, LOCIE, 73736 Le Bourget du Lac Cedex, France

Acoustic emission (AE) measurements performed during the compressive loading of concrete samples with three different microstructures (aggregate sizes and porosity) and four sample sizes revealed that failure is preceded by an acceleration of the rate of fracturing events, power law distributions of AE energies and durations near failure, and a divergence of the fracturing correlation length and time towards failure. This argues for an interpretation of compressive failure of disordered materials as a critical transition between an intact and a failed state. The associated critical exponents were found to be independent of sample size and microstructural disorder and close to mean-field depinning values. Although compressive failure differs from classical depinning in several respects, including the nature of the elastic redistribution kernel, an analogy between the two processes allows deriving (finite)-sizing effects on strength that match our extensive dataset. This critical interpretation of failure may have also important consequences in terms of natural hazards forecasting, such as volcanic eruptions, landslides, or cliff collapses.

Classical fracture and failure theoretical frameworks or criteria, such as Griffith theory or the Coulomb failure criterion, do not consider material disorder. Consequently, they predict an abrupt global failure, without any precursory phenomenon. In that sense, failure can be interpreted as a first-order transition from an intact to a failed state, as Griffith theory was inspired by the classical theory of nucleation [1,2]. Materials heterogeneity has been however considered for a long time, especially to account for failure strength variability and associated size effects [3]. Nevertheless, this weakest-link approach is based on strong assumptions such as the absence of mechanical interactions between defects and between rupture events, or a global failure dictated by the activation of the largest flaw (the weakest-link). These assumptions might appear reasonable for weakly disordered materials under tension, especially in the case of a pre-existing large crack or notch. However, in case of large enough disorder, the quasi-static propagation of such a crack can be interpreted as a dynamical critical transition [4,5]. The limitations of these classical frameworks appear even clearer for highly disordered systems without macro-scale heterogeneities [6] and/or loading conditions stabilizing crack propagation, such as compression (through the presence of friction). In those cases, it is known for a long time that failure is a *process*, involving the nucleation, interaction, propagation and coalescence of many microcracks [7,8], hence characterized by precursory phenomena. The presence/absence of precursors to failure and faulting has obvious consequences in terms of natural hazards forecasting, for e.g. earthquakes [9,10], cliff collapses [11], landslides [12], or volcanic eruptions [13,14].

The failure of heterogeneous media has been extensively studied over the last 30 years [1,15], essentially on the basis of theoretical and numerical models such as fiber-bundle (FBM) [16], random-fuse (RFM), random-spring (RSM) [17], or progressive damage (PDM) [2,18,19] models. However, the nature of the associated transition remains controversial. In the limit of infinite disorder, fracture can be mapped onto the percolation problem [20]. For bounded disorder, FBM with equal-load sharing, corresponding to a mean-field approximation, exhibit a critical behavior with the rate of bundle breaking per increasing stress diverging at the critical point (the failure) [16]. A critical transition was also reported for a PDM of compressive faulting, with the average damage avalanche size, the correlation length of damage [19], or the largest damage cluster [2], all progressively increasing during the loading history and diverging at failure. This interpretation is also consistent with a mapping of the faulting problem onto the depinning transition [21]. On the other hand, for RFM and RSM with large (but finite) disorder, it has been claimed that there is no diverging correlation length at failure [17,22], consistent with a first-order transition interpretation of failure in those models [23]. Besides the nature of the transition, this raises the question of the role of the disorder strength on failure precursors.

This debate calls for experimental data, which are still sparse and disparate. Power law distributions of acoustic emission (AE) energies released by damage and microcracking, $P(E) \sim E^{-\beta}$, have been frequently reported and presented as evidences of "criticality" in a broad sense. For highly porous [24,25] or cellular [26] materials under compression, the AE event rate $dN/dt$ or the energy distribution do not exhibit significant trends as approaching failure, possibly as the result of a transient hardening mechanism[27], whereas the (stable) power law pdf of energies is accompanied by Omori-like aftershocks triggering. However, other authors reported an acceleration of the event rate (time-reversed Omori's scaling) towards failure for materials with a porosity larger than 0.3, but an exponential growth (hardly compatible with the critical point hypothesis) for lower porosities[28,29]. In low-porosity rocks, a progressive localization of damage before faulting under compression has been revealed from either AE [7,30] or X-ray tomography [31,32]. In this last case, the damage rate, defined as the rate of increasing crack-induced porosity, as well as the size of the largest microcrack, were found to power-law diverge as approaching global failure, arguing for an interpretation of compressive faulting as a critical transition [32]. Criticality was also argued for the flexural failure of composite materials from a divergence of the AE energy release [33].

Hence, despite these various hints, experimental evidences are still lacking to ascertain this critical interpretation of failure, to determine the critical exponents, to check their universal character and to precise the role of internal disorder. To do so, we performed compression tests and AE measurements on an emblematic quasi-brittle heterogeneous material, concrete. Cylindrical samples with a constant aspect ratio ($L/D$ =2) but four different sizes ($L$=80, 140, 220, 320 mm) were prepared following French standards [34] and from three different concrete mixtures based on different aggregate sizes (Fine (F; i.e. only sand), Medium (M) and Coarse (C)). Disorder consisted of aggregates, sand particles, and pores, with a larger porosity for F-concrete ($\phi$ =4.8%) than for M- (1.6%) and C-concrete (1.5%). The microstructures as well as the elastic properties were sample size-independent, indicating that even the smallest samples were larger than the representative volume elements (RVE) of the materials [35]. The preparation procedure and the microstructural characterization of our materials have been detailed elsewhere [35].

Uniaxial compression was applied on each sample at a constant stress rate of 0.5 MPa/s, corresponding to a strain-rate between $2.4 \times 10^{-5}$ and $3.2 \times 10^{-5}$ s$^{-1}$. Loading was automatically stopped upon catastrophic failure, when the load dropped below 50% of peak load. Two (for $L$=80mm samples) to four (for $L$=140, 220 and 320mm samples) piezoelectric AE sensors with a frequency bandwidth of 20-1200 kHz were coupled directly to the samples sides using a silicon paste, and their signals pre-amplified at 40 dB. A standard procedure was used to detect AE bursts over a 30 dB amplitude threshold, and their characteristics (maximum amplitude $V_{max}$, energy $E$, duration $T$, ..) saved. A scaling analysis between $V_{max}$ and $T$ [36] indicates that the recorded voltage $V(t)$ is a good proxy of the seismic moment release rate, i.e. $T$ is a reasonable estimate of the duration of the fracturing event, for timescales larger than ~ 100 μs. This allowed tracking the fracturing process up to macroscopic failure, materialized by the development of an inclined fault throughout the sample. For each material, four tests were performed for $L$=80 mm samples, and two tests for $L$=140, 220 and 320 mm samples. The critical exponent values reported below result from an averaging over all these tests and all sensors.

We define the reduced control parameter as $\Delta = (\sigma_f - \sigma)/\sigma_f$, where $\sigma_f$ is the peak/failure stress, and conjecture that failure ($\Delta$=0) is a critical point. Fig. 1 show the evolution of the intermittent AE activity during a typical test, where the event rate, the total energy release and the maximum energy of events accelerate towards failure. Fig. 2 shows the AE event rate diverging towards failure following $dN/d\Delta \sim \Delta^{-p}$ with a maximum likelihood estimate[14] of $p$=2/3±0.05, independently of sample size (Fig. 2a) and material disorder (Fig. 2b; see also [36]). Such time-reversed Omori's law [37] has been reported for the compressive failure of various porous materials (14%≤$\phi$≤40%) [27-30,38], though with a varying $p$-value, possibly depending on the strain-rate [38]. In our low-porosity but disordered quasi-brittle materials, under our stress-controlled protocol, $p$ was found to be independent of both external and internal (disorder-related) scales.

This is accompanied by a progressive evolution of the distribution of AE energies as approaching failure (Fig. 3a). In the early stages of loading, the energy cumulative distribution (cdf), $P(>E)$, is clearly truncated towards large energies, but the associated upper cut-off is increasing as fracturing goes on. Close to failure ($\Delta \rightarrow 0$), a power law cdf is recovered, $P(>E) \sim E^{-\beta+1}$, over ~ 5 orders of magnitude, without detectable upper cut-off (Fig. 3a). We therefore conjecture an evolution of the probability density function (pdf), $P(E) \sim E^{-\beta_E} f(E/E^*)$, where $f(x)$ rapidly vanishes for $x>1$, and a cut-off energy diverging at the critical point, $E^* \sim \Delta^{-\gamma_E}$. From this, we recover a non-truncated power law distribution at failure, $P(E) \sim E^{-\beta_E}$, while the sweeping of an instability [39] predicts another power law $P(E) \sim E^{-\theta_E}$ with $\theta_E = \beta_E + 1/\gamma_E$ for the stress-integrated pdf (Fig. 3b). Our results support this conjecture with the exponents $\beta_E$ =1.4±0.06 and $\theta_E$ =1.75±0.04 not varying significantly with the sample size or the disorder. For each dataset (1 sensor on 1 sample), the exponents were determined from a maximum likelihood methodology [40]. This yields $\gamma_E = 1/(\theta_E - \beta_E)$ =3.3±0.5, a result that can be confirmed from a data collapse analysis (Fig. 3a). Combined with inverse-Omori acceleration, this evolution of energy distributions, which itself means an increase of the average energy $\langle E \rangle$ towards failure, implies a divergence of the energy release rate, $dE/d\Delta \sim \Delta^{-\alpha}$. We observed such evolution, independently of both sample size and disorder, however with an exponent $\alpha$=1.3±0.1 smaller than expected from a simple analysis [36]. Note that accelerations of the event rate and the energy release rate have been observed during the compressive failure of highly porous materials, although the energy distributions, and so $\langle E \rangle$, remained unchanged in this case. This indicates that these features are more generic than critical failure [27].

To translate this evolution of AE energies in terms of fracture size and correlation length, we consider an elastic crack model whose underlying hypotheses are (i) a compact (non-fractal) incremental crack/fault area $A$, (ii) an average slip or displacement proportional to the crack/fault "radius", $\langle u \rangle \sim r \sim A^{1/2}$, (iii) a $r$-independent stress drop, and (iv) a constant scaled energy, i.e. a radiated acoustic/seismic energy simply proportional to the potency $P_0 \sim \langle u \rangle A$ (or the seismic moment if multiplied by an elastic modulus). Such models are classical in AE analysis or seismology for both mode I cracks [41] or shear faults [42,43], well supported by available data [36], and lead to $E \sim r^3$ for the radiated acoustic energy. However, they differ from a depinning model of a planar fault [21,44] where the average slip $\langle u \rangle$ is independent of $r$, and ruptures can be fractal with $d_f \leq 2$, hence $P_0 \sim r^{d_f}$

. However, mean-field depinning also predicts a non-proportional scaling between the energy and the potency (or "size") of the avalanche, $E \sim P_0^{3/2}$ [45,46]. Taking the limiting case $d_f \approx 2$, this gives $E \sim r^3$ as for elastic crack models, though from a subtly different framework. This scaling yields for the cut-off incremental rupture radius, $r^{*3} \sim E^* \sim \Delta^{-\gamma_E}$. Further identifying $r^*$ with the correlation length $\xi$ of the fracturing/faulting process, one gets a divergence $\xi \sim \Delta^{-\nu}$ with $\nu = \gamma_E/3 = 1.1 \pm 0.2$ for $\Delta \to 0$.

Alike, an analysis of duration distributions during loading argues for a similar scaling, $P(T) \sim T^{-\beta_T} g(T/T^*)$, with a cut-off duration diverging at the critical point, $T^* \sim \Delta^{-\gamma_T}$, and exponents $\beta_T = 2.0 \pm 0.15$, $\theta_T = 2.9 \pm 0.1$ and $\gamma_T = 1.1 \pm 0.3$ independent of disorder and sample size (Fig. 3c). Although the uncertainty on these duration exponents is larger than for the energy distributions, this suggests $z = \gamma_T/\nu \approx 1$ for the dynamic exponent of the critical transition.

The power law distribution of microseismic energies and durations near failure, as well as the divergence of the rate of fracturing events, of the fracturing correlation length, and of the associated duration as approaching failure, are strong evidences for an interpretation of the compressive failure of low-porosity disordered materials as a critical transition, where the failure stress identifies as the critical point. This is further supported by the independence of the critical exponents relatively to sample size and disorder. Theoretically, this critical interpretation could be checked from a finite-size scaling analysis of energy and duration pdfs obtained from samples of different sizes. We did not find, however, a fully convincing $L$-dependence in our data, most likely because (i) the size range explored was limited ($L_{max}/L_{min} = 4$) and (ii) the necessarily limited experimental data statistics make the analysis of extremes difficult.

From these results, the nature of the critical transition, and its possible affiliation to a particular universality class, can be further discussed. A mapping of the problem of stick-slip along an existing fault to the depinning of an elastic interface was proposed 20 years ago [21,44]. More recently, a similar analogy was proposed in case of compressive failure to account for statistical size effects on strength [47] (see below). Indeed, quasi-brittle failure shares fundamental ingredients with the depinning transition, including a local threshold mechanism, disorder, and elastic interactions. Our results reveal a similar phenomenology of avalanches as approaching the critical point, with experimental exponents remarkably close to mean-field depinning ones [45,48,49] (see [36]). On the other hand, several differences between the two problems can be stressed. First, the time-reversed Omori's scaling of the avalanche rate is not present in classical depinning, meaning that an additional exponent, $p$, is required to describe the failure transition. In addition, the nature of the elastic interaction kernel differs. Unlike for depinning, it is non-convex in our case [47,50], allowing localization of damage along a fault, much like for the yielding transition in amorphous plasticity [51,52]. It is non-negative either, meaning that is has unstable modes, differing on this point from the yielding transition [50]. Although these differences preclude a direct affiliation of our problem onto the universality class of classical depinning, the scaling of the fracturing correlation length, $\xi \sim \Delta^{-\nu}$, with an exponent very close to mean-field depinning ($\nu_{MF} = 1$ [48]), suggests that some theoretical results could be tentatively transposed to our problem. In particular, we can expect a finite-size effect on failure stress, yielding the following scaling laws for both the mean

$$\langle \sigma_f \rangle = \sigma_\infty (L/L_m)^{-1/\nu_{FS}} + \sigma_\infty \quad (1)$$

, and the standard deviation

$$\delta(\sigma_f) = \sigma_\infty (L/L_\delta)^{-1/\nu_{FS}} \quad (2)$$

, where $\sigma_\infty$ is the asymptotic strength of infinitely large systems, and $L_m$ and $L_\delta$ are length scales related to the microstructural disorder characteristic length scale[47]. The classical assumption is $\nu_{FS} = \nu$ [53], allowing to relate the critical nature of the failure process to statistical size effects on strength. The above predictions were recently confirmed from an extensive series (527) of similar compression tests on the same materials and the same sample sizes (though without AE monitoring), using a mean-field depinning prediction for the exponent, $\nu_{FS} = 1$ [35]. Here we re-analyzed these data using the AE-derived exponent $\nu_{FS} = \nu = 1.1$. While the length scales $L_m$ and $L_\delta$ (obtained from best-fitting of the above equations with the data, taking $\nu_{FS} = 1.1$) strongly depend on the material, and particularly its pore structure, the asymptotic strength $\sigma_\infty$ was found to be essentially independent of the microstructural disorder. Consequently, the accuracy and the universal character of Eqs. (1) and (2) to accounts for experimental size effects on compressive strength of quasi-brittle materials can be demonstrated on a rescaled plot (Fig. 4). Beyond an independent confirmation of the pertinence of the theoretical framework, these results strikingly illustrate the usefulness of statistical physics theoretical concepts accounting for the mechanical behavior of disordered quasi-brittle materials and, in the end, to constrain engineering regulations [35].

The present work is also important in the context of a possible forecast of geophysical hazards. Time-reversed Omori's scaling has been proposed to forecast volcanic eruptions from seismic data[54], though with a large forecast uncertainty inherent in the form of this law (rate diverging towards the critical point)[14]. An evolution of the energy distribution of seismic signals, similar to that reported in Fig. 3a, has been reported within 2 hours before a chalk cliff collapse [11], but additional analysis would be needed to precisely checked this analogy in terms of mechanisms and critical exponents. The possible prediction of large, devastating earthquakes is a long-standing, still unsolved problem. Many large earthquakes seem to be preceded by foreshocks, and a time-reversed Omori's law as well as a divergence of the seismic moment release rate have been sometimes reported [10]. However, these precursory phenomena are far to be ubiquitous [9,55], and foreshocks could actually be just an expression of cascades of triggered seismicity implying that earthquakes are "predictable" to the same degree whatever their size[56]. Hence, the use of these potential precursors as a forecasting tool remains elusive. This raises fundamental questions, such as the difference between the compressive failure of initially un-faulted rocks [31,32] and the earthquake nucleation along a pre-existing crustal fault, and calls for further theoretical and experimental work as well as geophysical data analysis.

## ACKNOWLEDGEMENTS

This work has been supported by the AGIR program of Université Grenoble-Alpes. Anonymous referees are thanked for interesting comments and suggestions .

* jerome.weiss@univ-grenoble-alpes.fr

[1] M. J. Alava, P. K. V. V. Nukala, and S. Zapperi, Advances in Physics **55**, 349 (2006).
[2] L. Girard, J. Weiss, and D. Amitrano, Phys. Rev. Lett. **108**, 225502 (2012).
[3] W. Weibull, Proc. Royal Swedish Academy of Eng. Sci. **151**, 1 (1939).
[4] D. Bonamy, S. Santucci, and L. Ponson, Phys. Rev. Lett. **101**, 045501 (2008).
[5] J. Barés, L. Barbier, and D. Bonamy, Phys. Rev. Lett. **111**, 054301 (2013).
[6] J. Davidsen, G. Kwiatek, E.-M. Charalampidou, T. Goebel, S. Stanchits, M. Rück, and G. Dresen, Phys. Rev. Lett. **119**, 068501 (2017).
[7] D. A. Lockner, J. D. Byerlee, V. Kuksenko, A. Ponomarev, and A. Sidorin, Nature **350**, 39 (1991).
[8] Z. e. Reches and D. A. Lockner, Journal of Geophysical Research: Solid Earth **99**, 18159 (1994).
[9] L. de Arcangelis, C. Godano, J. R. Grasso, and E. Lippiello, Physics Reports **628**, 1 (2016).
[10] S. C. Jaumé and L. R. Sykes, PAGEOPH **155**, 279 (1999).
[11] D. Amitrano, J. R. Grasso, and G. Senfaute, Geophysical Research Letters **32** (2005).
[12] G. B. Crosta and F. Agliardi, Canadian Geotechnical Journal **40**, 176 (2003).
[13] R. Sparks, Earth and Planetary Science Letters **210**, 1 (2003).
[14] A. F. Bell, M. Naylor, and I. G. Main, Geophysical Journal International **194**, 1541 (2013).
[15] H. J. Herrmann and S. Roux, *Statistical models for the fracture of disordered media* (North-Holland, Amsterdam, 1990).
[16] S. Pradhan, A. Hansen, and B. K. Chakrabarti, Rev. Mod. Phys. **82**, 499 (2010).
[17] P. K. V. V. Nukala, S. Zapperi, and S. Simunovic, Phys. Rev. E **71**, 066106 (2005).
[18] D. Amitrano, J. R. Grasso, and D. Hantz, Geophys. Res. Lett. **26**, 2109 (1999).
[19] L. Girard, D. Amitrano, and J. Weiss, J. Stat. Mech. **P01013** (2010).
[20] S. Roux, A. Hansen, H. Herrmann, and E. Guyon, J. Stat. Phys. **52**, 237 (1988).
[21] D. S. Fisher, K. Dahmen, S. Ramanathan, and Y. Ben-Zion, Phys. Rev. Lett. **78**, 4885 (1997).
[22] P. Nukala, S. Simunovic, and S. Zapperi, J. Stat. Mech.-Theory Exp., P08001 (2004).
[23] S. Zapperi, P. Ray, H. E. Stanley, and A. Vespignani, Phys. Rev. Lett. **78**, 1408 (1997).
[24] J. Baró, Á. Corral, X. Illa, A. Planes, E. K. Salje, W. Schranz, D. E. Soto-Parra, and E. Vives, Phys. Rev. Lett. **110**, 088702 (2013).
[25] G. F. Nataf, P. O. Castillo-Villa, P. Sellappan, W. M. Kriven, E. Vives, A. Planes, and E. K. Salje, Journal of Physics: Condensed Matter **26**, 275401 (2014).
[26] T. Mäkinen, A. Miksic, M. Ovaska, and M. J. Alava, Phys. Rev. Lett. **115**, 055501 (2015).
[27] J. Baró, K. A. Dahmen, J. Davidsen, A. Planes, P. O. Castillo, G. F. Nataf, E. K. Salje, and E. Vives, Phys. Rev. Lett. **120**, 245501 (2018).
[28] J. Vasseur, F. B. Wadsworth, Y. Lavallée, A. F. Bell, I. G. Main, and D. B. Dingwell, Scientific reports **5**, 13259 (2015).
[29] J. Vasseur, F. B. Wadsworth, M. J. Heap, I. G. Main, Y. Lavallée, and D. B. Dingwell, Earth and Planetary Science Letters **475**, 181 (2017).
[30] A. Schubnel, B. Thompson, J. Fortin, Y. Guéguen, and R. Young, Geophysical research letters **34** (2007).
[31] F. Renard, B. Cordonnier, M. Kobchenko, N. Kandula, J. Weiss, and W. Zhu, Earth Planet. Sci. Lett. **476**, 69 (2017).
[32] F. Renard, J. Weiss, J. Mathiesen, Y. B. Zion, N. Kandula, and B. Cordonnier, Journal of Geophysical Research: Solid Earth **123**, JB014964 (2018).
[33] A. Guarino, A. Garcimartin, and S. Ciliberto, Eur. Phys. J. B. **6**, 13 (1998).
[34] NF-EN, edited by A. F. d. N. (AFNOR)2004).
[35] C. C. Vu, J. Weiss, O. Plé, D. Amitrano, and D. Vandembroucq, J. Mech. Phys. Solids **121**, 47 (2018).
[36] See Supplemental Material [url], which includes Refs. [57-65], for (i) a description of mechanical tests and acoustic emission recording, (ii) an interpretation of AE durations, (iii) a table of critical exponents, (iv) details about the maximum likelihood estimation of the $p$-value, (v) observational support for the crack/fault source model, (vi) details about AE energy release rate, and (vii) a brief discussion about the AE energy distributions.
[37] T. Utsu, Y. Ogata, and R. Matsu'ura, Journal of Physics of the Earth **43**, 1 (1995).
[38] I. O. Ojala, I. G. Main, and B. T. Ngwenya, Geophysical research letters **31** (2004).
[39] D. Sornette, Journal de Physique I **4**, 209 (1994).
[40] A. Clauset, C. R. Shalizi, and M. E. J. Newman, SIAM Rev. **51**, 661 (2009).
[41] A. G. Evans, in *Fundamentals of acoustic emission*, edited by K. Ono (University of California, Los Angeles, 1979), pp. 209.
[42] J. D. Eshelby, Proc. Roy. Soc. A **241**, 376 (1957).
[43] P. M. Shearer, *Introduction to seismology* (Cambridge University Press, 2009).
[44] D. S. Fisher, Physics Reports **301**, 113 (1998).

[45] E. K. Salje and K. A. Dahmen, Annu. Rev. Condens. Matter Phys. **5**, 233 (2014).
[46] K. A. Dahmen, in *Avalanches in Functional Materials and Geophysics* (Springer, 2017), pp. 19.
[47] J. Weiss, L. Girard, F. Gimbert, D. Amitrano, and D. Vandembroucq, PNAS **111**, 6231 (2014).
[48] D. Ertas and M. Kardar, Phys. Rev. E **49**, R2532 (1994).
[49] M. LeBlanc, L. Angheluta, K. Dahmen, and N. Goldenfeld, Phys. Rev. E **87**, 022126 (2013).
[50] V. Démery, V. Dansereau, E. Berthier, L. Ponson, and J. Weiss, arXiv preprint arXiv:1712.08537 (2017).
[51] J. Lin, E. Lerner, A. Rosso, and M. Wyart, PNAS **111**, 14382 (2014).
[52] B. Tyukodi, S. Patinet, S. Roux, and D. Vandembroucq, Phys. Rev. E **93**, 063005 (2016).
[53] O. Narayan and D. S. Fisher, Phys. Rev. B **48**, 7030 (1993).
[54] I. Main, Geophys. J. Int. **139**, F1 (1999).
[55] M. Bouchon, V. Durand, D. Marsan, H. Karabulut, and J. Schmittbuhl, Nat. Geosci. **6**, 299 (2013).
[56] A. Helmstetter and D. Sornette, Journal of Geophysical Research: Solid Earth **108** (2003).
[57] S. Deschanel, W. Ben Rhouma, and J. Weiss, Scientific Reports **7**, 13680 (2017).
[58] C. H. Scholz and P. A. Cowie, Nature **346**, 837 (1990).
[59] J. Walsh, J. Watterson, and G. Yielding, Nature **351**, 391 (1991).
[60] S. Goodfellow and R. Young, Geophysical Research Letters **41**, 3422 (2014).
[61] E. J. Sellers, M. O. Kataka, and L. M. Linzer, J. Geophys. Res. **108**, doi:10.1029/2001JB000670 (2003).
[62] H. Kanamori and E. E. Brodsky, Reports on Progress in Physics **67**, 1429 (2004).
[63] I. G. Main, P. G. Meredith, and C. Jones, Geophysical Journal International **96**, 131 (1989).
[64] X. Jiang, H. Liu, I. G. Main, and E. K. Salje, Phys. Rev. E **96**, 023004 (2017).
[65] D. Amitrano, The European Physical Journal Special Topics **205**, 199 (2012).

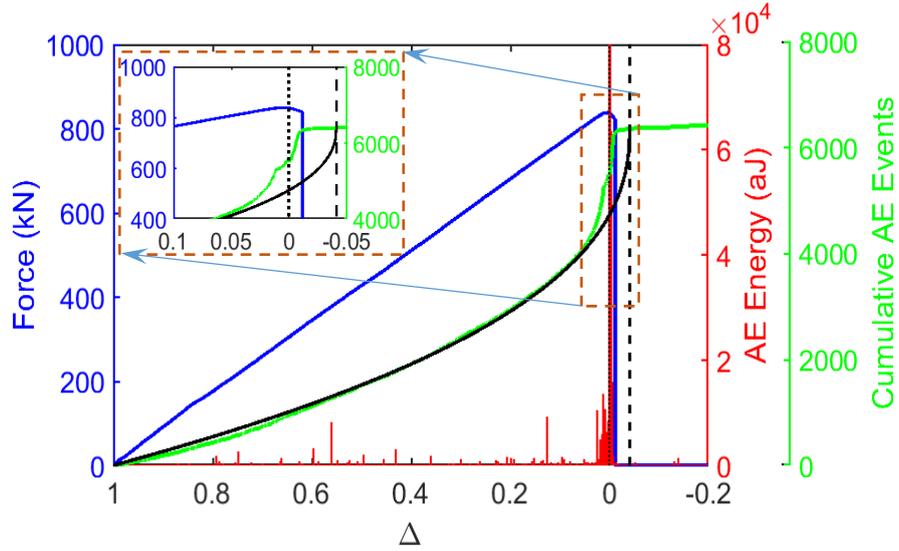

FIG 1. Evolution of the AE activity during a stress-controlled compression test on a sample of length *L*=160 mm of M-concrete. Blue curve: load; green curve: cumulated number of AE events; black curve: same as the green one for a theoretical time-reversed Omori's law with parameters estimated from a maximum likelihood method [36]; red: AE energy release rate, sampled at 100 Hz. The black dotted line represents the observed failure (maximum) stress, while the black dashed line represents the failure stress predicted from the theoretical time-reversed Omori's law.

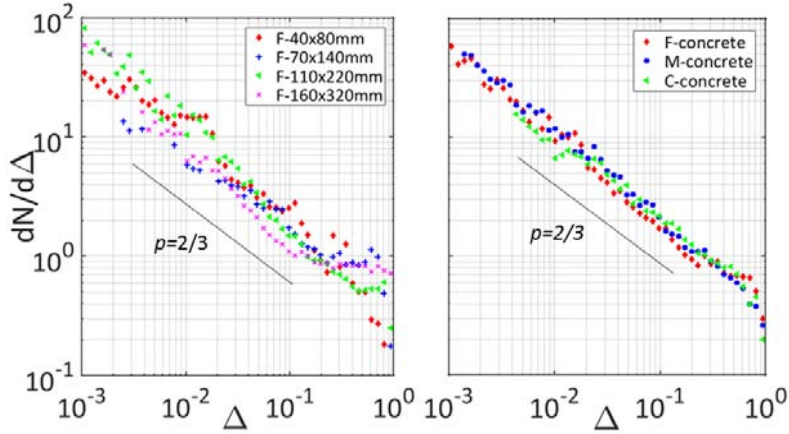

FIG. 2. AE event rate $dN/d\Delta$ (left) for different sample sizes of F-concrete, and (right) for the three different materials. Curves on the left were averaged over all sensors and all samples of a given size, and on the right on all sensors and samples (whatever the size) of a given material.

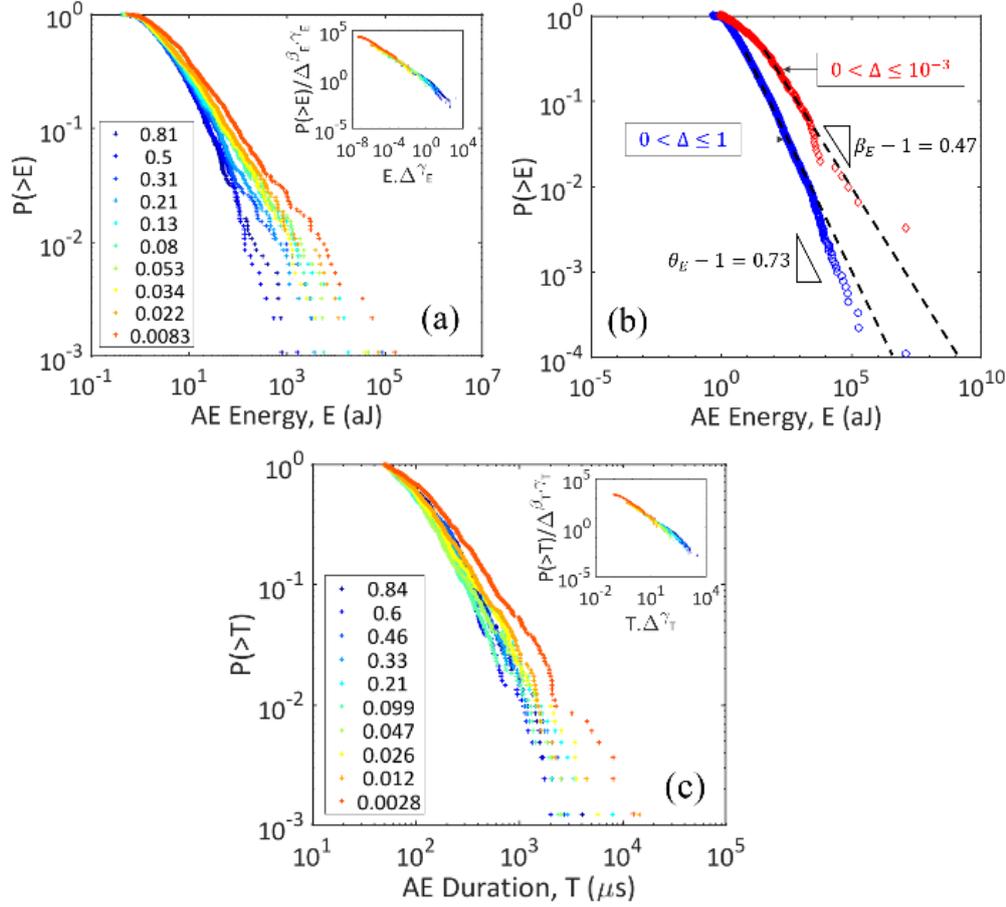

FIG. 3. Evolution of the distributions of AE energies and durations as approaching failure. (a) Cumulative distributions (cdf) of AE energies at different distances to failure $\Delta$, for a test on a 110 mm sample of M-concrete. Each distribution is built from at least 300 events. Inset: data collapse of the same data in a rescaled plot. Other sample sizes and materials give similar results. (b) Near failure (red diamonds), and stress-integrated (blue circles) cumulative distribution of AE energies for a test on a 40 mm sample of F-concrete. Other sample sizes and materials give similar results. (c) Same as (a) for the cumulative distributions of AE durations above 50 μs. At smaller timescales, the measured AE durations, influenced by wave scattering and seismic coda, are not a good proxy of avalanche durations [36].

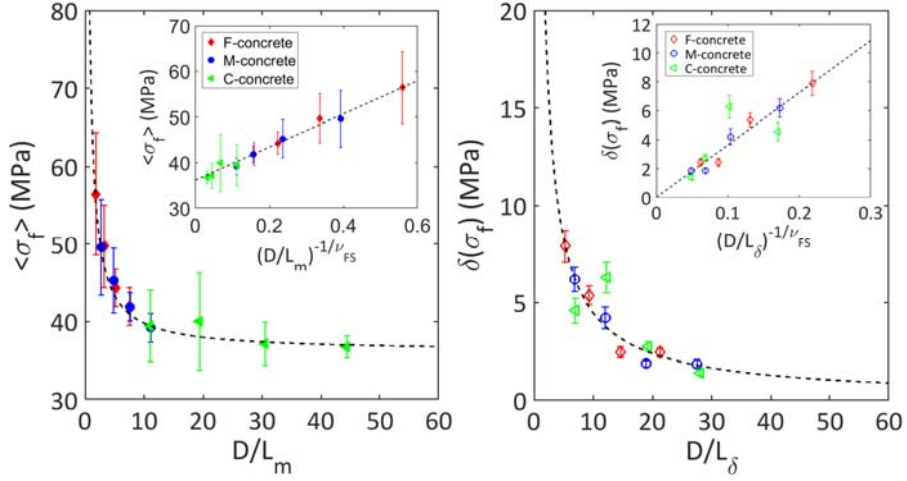

FIG. 4. Size effects on the average strength (left) and the associated standard deviation (right) of three types of concrete from a statistical dataset of 527 compression tests. In these plots, the external sample diameter ($D$) is normalized by the internal (disorder-related) scales $L_m$ (left) and $L_\delta$ (right), and the asymptotic, material-independent strength is $\sigma_\infty$=36.2 MPa. The dotted lines represent the finite-size scaling predictions with $\nu_{FS}=1.1$ (see text for details). Insets: same data in rescaled plots to show the asymptotic strength, or the vanishing fluctuations towards the thermodynamic limit ($\lim_{L\to\infty}\delta(\sigma_f)=0$).

# Compressive failure as a critical transition:
# Experimental evidences and mapping onto the universality class of depinning
# -
# Supplemental Material

Chi-Cong Vu[1], David Amitrano[1], Olivier Plé[2] and Jérôme Weiss[1,*]

[1]Univ. Grenoble Alpes, CNRS, ISTerre, 38000 Grenoble, France
[2]Univ. Savoie Mont-Blanc, CNRS, LOCIE, 73736 Le Bourget du Lac Cedex, France

I. Mechanical experiments and acoustic emission recording

The preparation of the concrete samples as well as the mechanical experimental setup have been extensively described elsewhere [35]. The microstructure of the three concretes is illustrated on Fig. S1(a). Microstructural characteristics were obtained from image analyses of internal sections such as those of Fig. S1(a) [35]. They are summarized below.

| Concrete Mixture | Correlation length of the global microstructure (mm) | Correlation length of the pore microstructure (μm) | Mean pore diameter (mm) | Maximum pore diameter (mm) | Porosity |
|---|---|---|---|---|---|
| Fine | 0.6 | 26.4 | 0.33 | 6.9 | 0.048 |
| Medium | 2.1 | 9.8 | 0.31 | 6.7 | 0.016 |
| Coarse | 3.5 | 8.5 | 0.28 | 5.4 | 0.015 |

Table S1. Main microstructural characteristics of the three concrete materials.

For each concrete, Acoustic Emission (AE) was recorded during compression tests performed on four samples sizes: 4 tests for $L$=80 mm samples, and 2 tests for $L$=140, 220 and 320 mm samples (Fig. S1(b)). This represented a total of 30 mechanical tests with AE recording (Fig. S1(c)). In addition, we show on Fig. 4 of the main text the size effects on strength for the same materials obtained from an extensive campaign of 527 mechanical tests performed without AE recording, and analyzed elsewhere in details [35].

In this study, resonant AE sensors of type PICO produced by Physical Acoustics Corporation were used. Their frequency range is about 20 kHz-1.2 MHz, with a peak frequency of approximately 900 kHz. Their small size (4×5 mm) make them easy to couple on our small samples (i.e. $L$=80-mm samples). We used a Silicone adhesive glue (Silcoset 151) for coupling. In order to ensure a proper coupling, some small areas on the lateral surfaces of the samples were ground and polished by an angle grinder with less and less grit size of metal-bonded discs. This ensured that the sample surface were not damaged. The AE signals from the loaded specimen are converted into electrical signals by the AE sensors, then pre-amplified with a gain of 40 dB and recorded by the Acoustic Emission Digital Signal Processor (AEDSP-32/16) cards at a sampling rate of 4 MHz. To detect AE bursts, we used an amplitude threshold of 30 dB, a Peak Definition Time (PDT) of 10 µs, a Hit Definition Time (HDT) of 20 µs and a Hit Lock-out Time (HLT) of 20 µs. These parameters were defined by performing AE recording on samples before loading but with the loading machine switched on (to set the AE threshold relatively to the environmental noise amplitude), as well as Pencil Lead Break tests which are similar to Hsu-Nielsen tests (to set PDT, HDT and HLT),.

The burst duration $T$ is defined as the time over which the envelope of the AE signal $V(t)$ remains above the threshold $V_{th}$=30 dB, while the energy $E$ is calculated from the integral of the squared voltage divided by the reference resistance, over the duration of the AE waveform, $E \sim \int_T V(t)^2 dt$.

In addition to this discrete AE, we also continuously recorded the AE power $dE/dt$ at a sampling rate of 100 Hz.

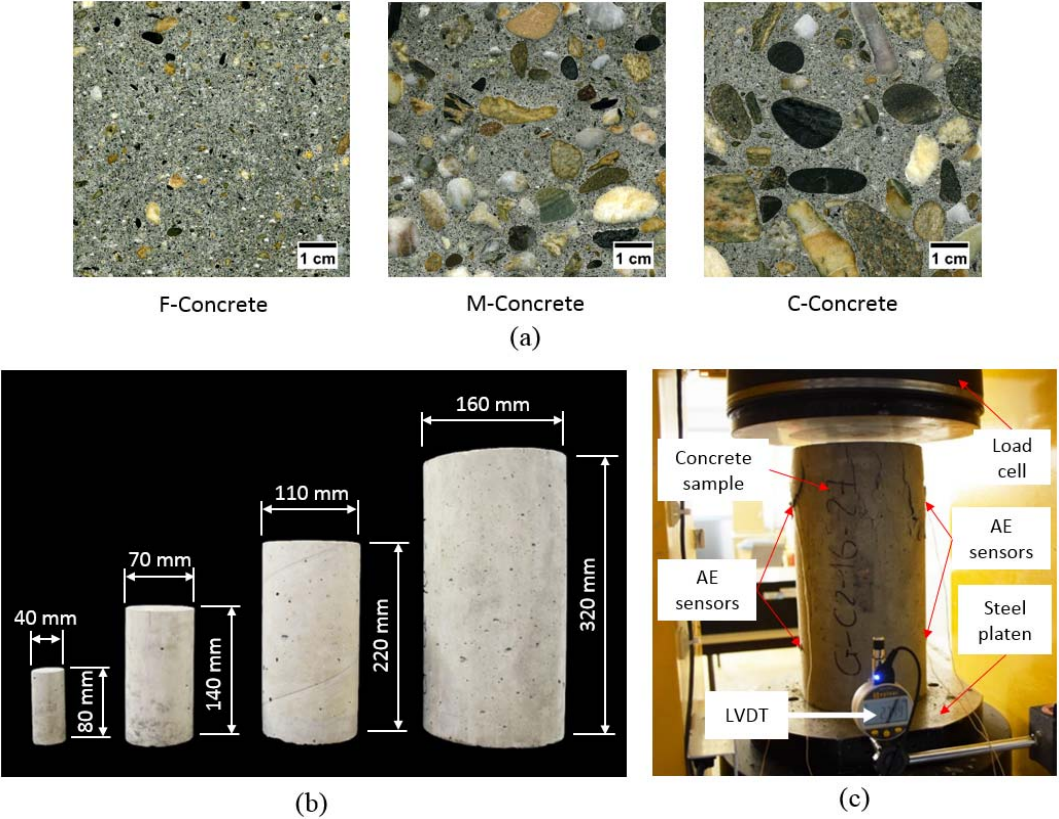

FIG. S1. Concrete samples used for the experimental investigations: (a) Cross sections of the three different concrete mixtures; (b) Geometries of the four different sizes of concrete samples; (c) Experimental set-up with AE recording.

## II. Interpretation of AE durations

The interpretation of the AE duration $T$ in terms of duration of the fracturing event itself is not always straightforward (e.g.[27]). Indeed, the duration is typically defined as the time over which the envelope of the AE signal $V(t)$ remains above a chosen threshold $V_{th}$. As the result of scattering of the wave generated at the source by internal disorder and/or reflections at free surface, a coda can develop after the initial pulse (e.g. [57]). If the material damping is low, this coda will strongly influence the measured duration of the event, which hence will lose its physical meaning in terms of genuine duration of the source mechanism. In this case, we expect an exponential decay for the amplitude of the signal, $V(t) = V_{max}\exp((t_0 - t)/\tau)$, where $\tau$ is the attenuation time scale, essentially dictated by material properties, and $t_0$ the event arrival time. Assuming a short rise time, i.e. $(t = t_0) \approx V_{max}$, this yields:

$$T = \tau(\log(V_{max}) - \log(V_{th})) \tag{S1}$$

, or:

$$\langle V_{max}|T\rangle = V_{th}\exp(T/\tau) \tag{S2}$$

for the conditional average $\langle V_{max}|T\rangle$. Therefore, $\langle V_{max}\rangle \approx V_{th}$ for $T \ll \tau$, and should grow exponentially for $T > \tau$. The first prediction is recovered for durations below ~100 μs, but another scaling is observed above, $\langle V_{max}|T\rangle \sim T^\delta$, with $\delta$=0.95±0.05 (Fig. S2). This argues for an attenuation timescale of about 100 μs, and so, for larger timescales, the voltage signal $V(t)$ is a good proxy of the seismic moment release rate, or, in other words, of the avalanche velocity $v(t)$ [27]. Note that the measured value of $\delta$ is in good agreement with mean-field depinning ($\delta$=1; [49]). The increasing scatter observed at large durations likely comes from (i) poorer statistics for large events and (ii) an increasing probability to merge few successive events into an apparently single one when the average AE activity becomes very large near failure.

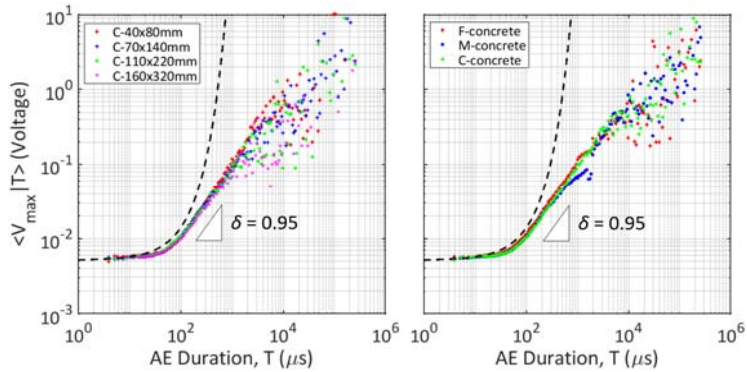

FIG. S2. Conditional average maximum AE amplitude $\langle V_{max}|T\rangle$ for (left) different sample sizes of C-concrete, and (right) the three different materials. Curves on the left were averaged over all sensors and all samples of a given size, and on the right on all sensors and samples (whatever the size) of a given material. The dotted black lines represent equation S2 with $\tau$ =100 μs.

III. Critical exponents and comparison with Mean-Field Depinning/Stick-Slip values

From the scaling analyses presented in the main text and in section I above, a full set of critical exponents can be derived. The values summarized in Table S1 have been averaged over all samples of a given material, whatever their size, as we did not find any significant size dependence, as expected for critical exponents. The experimental values are compared with mean-field depinning/stick-slip predictions. Overall, the agreement is remarkable. This validates the analogy between compressive failure and depinning proposed in [47]. Note however that the classical depinning framework does not predict an inverse Omori's law, $dN/d\Delta \sim \Delta^{-p}$, as observed in our data. This, in addition with an elastic redistribution kernel of a different nature [50], precludes an exact mapping of the failure problem onto classical depinning.

| Quantity | Form | Exponent | Concrete group ||| All groups | Mean-field values [45,48,49] |
|---|---|---|---|---|---|---|---|
| | | | F-concrete | M-concrete | C-concrete | | |
| Duration distribution | $P(T) \sim T^{-\beta_T} g(\Delta^{\gamma_T} T)$ | $\beta_T$ | 2.1 ± 0.2 | 2.0 ± 0.3 | 2,0 ± 0.3 | **2.0 ± 0.15** | **2** |
| | | $\gamma_T$ | 1.2 ± 0.5 | 1.1 ± 0.4 | 1.1 ± 0.5 | **1.1 ± 0.3** | **1** |
| Stress-integrated duration distribution | $P_{int}(T) \sim T^{-\theta_T}$ | $\theta_T$ | 2.9 ± 0.2 | 2.9 ± 0.2 | 2.9 ± 0.2 | **2.9 ± 0.1** | **3** |
| Energy distribution | $P(E) \sim E^{-\beta_E} f(\Delta^{\gamma_E} E)$ | $\beta_E$ | 1.5 ± 0.1 | 1.4 ± 0.1 | 1.4 ± 0.1 | **1.4 ± 0.06** | **4/3** |
| | | $\gamma_E$ | 3.2 ± 1.1 | 3.3 ± 0.8 | 3.4 ± 0.8 | **3.3 ± 0.5** | **3** |
| Stress-integrated energy distribution | $P_{int}(E) \sim E^{-\theta_E}$ | $\theta_E$ | 1.8 ± 0.1 | 1.75 ± 0.05 | 1.75 ± 0.05 | **1.75 ± 0.04** | **5/3** |
| Conditional average maximum amplitude vs. Duration | $\langle V_{max}|T \rangle \sim T^\delta$ | $\delta$ | 0.95 ± 0.05 | 0.90 ± 0.05 | 0.95 ± 0.05 | **0.95 ± 0.03** | **1** |
| Rate of AE event | $dN/d\Delta \sim \Delta^{-p}$ | $p$ | 0.64 ± 0.05 | 0.67 ± 0.09 | 0.66 ± 0.05 | **0.66 ± 0.05** | **None** |
| ***Derived exponents*** | | | | | | | |
| Correlation length | $\xi \sim \Delta^{-\nu}$ | $\nu$ | | | | **1.1 ± 0.2** | **1** |
| Dynamic exponent | $T^* \sim \xi^z$ | $z$ | | | | **1.0 ± 0.35** | **1** |

Table S2. Experimentally derived scaling laws and associated exponents for compressive failure. The experimental values are compared with mean-field depinning predictions [45,48,49].

## IV. Maximum likelihood estimation of the *p*-value

Figure 2 of the main text indicates a time-reverse Omori's acceleration of the event rate $\frac{dN}{d\Delta}$ towards failure, with a *p*-value independent of sample size and the material. To estimate the *p*-value and to check the accuracy of this acceleration up to failure, we performed a maximum likelihood analysis [14]. The time-reverse Omori's law is expressed as

$$\frac{dN}{d\Delta} = k \left( \frac{\sigma_f - \sigma}{\sigma_f} \right)^{-p} \quad (S3)$$

For $p \neq 1$, the log-likelihood function for this law for a catalogue of *n* events occurring at stresses $\sigma_{1 \leq i \leq n}$ within a stress interval $[\sigma_0, \sigma_e]$ writes [14]:

$$\log L(k, \sigma_p, p) = n \log k - p \sum_{i=1}^{n} \log(\sigma_p - \sigma_i) - \frac{k}{1-p} \left[ (\sigma_p - \sigma_e)^{1-p} - (\sigma_p - \sigma_0)^{1-p} \right] \quad (S4)$$

, where $\sigma_p$ is the predicted failure stress (at which $\frac{dN}{d\Delta}$ diverges).

For each sensor and each sample (corresponding to a single AE record), we searched for the set of parameters $(k, \sigma_p, p)$ maximizing this log-likelihood function, taking $\sigma_0 = 0.1 \sigma_f$ and $\sigma_e = 0.9 \sigma_f$. The conclusions of this analysis are:

(i) The maximum log-likelihood estimate of the *p*-value was found to be independent of sample size and the material, in agreement with Fig. 2 of the main text, and with an average $p = 2/3 \pm 0.05$.

(ii) For all materials and all sample sizes, the difference between the predicted, $\sigma_p$, and the measured, $\sigma_f$, failure stress was found, in average (for a given material and a given sample size, averaging performed over different AE sensors and specimens) to be less than 4% (see e.g. Fig. 1 of the main text).

(iii) For a given material and/or sample size, the difference between $\sigma_p$ and $\sigma_f$ could be either positive (failure predicted after observed failure), or negative (failure predicted before observed failure), although with a slight deficit of negative values. Overall, the mean value of $\sigma_p$ (for a given material and sample size) overestimates the observed failure stress by 3% at most.

(iv) The accuracy of the prediction did not depend significantly on the sample size or on the material (and so on the porosity).

## V. Observational support for the elastic crack/fault source model

$\langle u \rangle \sim r$ scaling: the elastic fault model is supported by a compilation of field measurements of mean fault slip against fault length [58], over a scale range $0.1 \, m \leq r \leq 1000 \, km$. Note that [59] argued for another scaling, $\langle u \rangle \sim r^{1.5}$, which is however lacking a clear theoretical/mechanical interpretation.

$E \sim r^3$: In seismology, the source radius can be estimated from the corner frequency of the spectrum, $f_c$: $r \sim 1/f_c$; Compiling in-situ micro-seismic (including some AE data) and seismic data, Goodfellow and Young [60] showed that (i) the seismic moment $M_0$ scales as $f_c^{-3}$, i.e. as $r^3$, for moments ranging from $10^{-2}$ to $10^{17}$ N.m; and (ii) the radiated energy is proportional to the seismic moment over the range $10^{-2}$ to $10^8$ N.m. These two results are also fully consistent with those reported by Sellers et al. [61] over scales ranging from lab-scale AE to earthquakes. Note that, in our case, the resonant character of our sensors makes such corner frequency analysis difficult. A combination of (i) and (ii) gives the expected scaling $E \sim r^3$. At larger moments, the scaling of the scaled energy is still highly debated nowadays, either $E \sim M_0$, or maybe $E \sim M_0^{5/4}$ (e.g. [43,62]). As mentioned by Sellers et al.[61], if some individual datasets over specific scale ranges (such as AE tests, or mine microseismicity) can suggest a non-linear scaling ($E \sim M_0^x$ with $x$ slightly >1), the combination of all datasets is fully consistent with $E \sim M_0$. Finally, Kanamori and Brodsky [62] reported observations also arguing for a scaling $M_0 \sim r^3$ for $1\ km \leq r \leq 100\ km$.

Taken altogether, these observations support the hypotheses of the elastic crack model used in our work.

## VI. AE energy release rate

From the combination of an inverse Omori law, $dN/d\Delta \sim \Delta^{-p}$, and a divergence of the energy cut-off, $E^* \sim \Delta^{-\gamma_E}$, one expects a power law divergence of the energy release rate, $dE/d\Delta$, as approaching failure. Using the AE power records sampled at 100 Hz (see above), we found such divergence,

$$\frac{dE}{d\Delta} \sim \Delta^{-\alpha} \tag{S5}$$

, with an exponent $\alpha=1.3\pm0.1$ independent of the sample size or the microstructural disorder, as expected (Fig. S3).

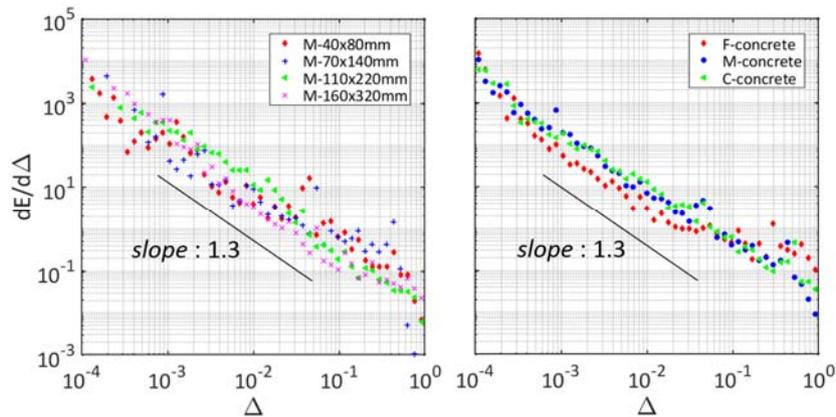

FIG. S3. The AE energy release rate $dE/d\Delta$ for (left) different sample sizes of M-concrete, and (right) for three different concrete mixtures. Curves on the left were averaged over all sensors and all samples of a given sample size, and on the right on all sensors and samples (whatever the sample size) of a given material. Other sample sizes for F- and C-concretes give similar results.

In principle, $\alpha$ can be related with the exponents $p$ and $\gamma_E$ from a simple analysis. Indeed,

$$\frac{dE}{d\Delta} = \frac{dE}{dN}\frac{dN}{d\Delta} = \langle E \rangle \frac{dN}{d\Delta} \tag{S6}$$

Then,

$$\langle E \rangle = \int_{E_m}^{+\infty} E.P(E)dE \tag{S7}$$

, where $E_m$ is the minimum AE Energy at a given $\Delta$ (lower cut-off). Using the form of the energy distribution, $P(E) \sim E^{-\beta_E} f(\Delta^{\gamma_E} E)$, one gets:

$$\langle E \rangle = \int_{E_m}^{+\infty} E^{1-\beta_E}.f\left(\frac{E}{E^*}\right) dE \tag{S8}$$

Assuming that $P(E)$ is brutally truncated at $E^*$, we obtain:

$$\langle E \rangle \sim \int_{E_m}^{E^*} E^{1-\beta_E} dE \tag{S9}$$

When approaching failure, the cut-off value $E^*$ is much larger than $E_m$, and the exponent $\beta_E$ is always larger than 1. Hence, we obtain:

$$\langle E \rangle_{(\Delta \to 0)} \sim (E^*)^{2-\beta_E} \left[1 - \left(\frac{E_m}{E^*}\right)^{2-\beta_E}\right] \tag{S10}$$

Close to failure where $E^* \gg E_m$, one thus should get:

$$\langle E \rangle_{(\Delta \to 0)} \sim (E^*)^{2-\beta_E} \tag{S11}$$

Combined with (S4) and the inverse Omori scaling, this gives

$$\frac{dE}{d\Delta} \sim \Delta^{-(\gamma_E(2-\beta_E)+p)} \tag{S12}$$

, i.e. $\alpha = \gamma_E(2 - \beta_E) + p$. If our data are consistent with a power law divergence of the AE power (Fig. S3), the measured exponents $\alpha$ were found to be significantly smaller than what we could expect from this simple derivation and the values of $p$ and $\gamma_E$ as, from table S2, $\gamma_E(2 - \beta_E) + p = 2.6$. The origin of this discrepancy is still partly obscure, and might be related either to the nature of experimental AE fluctuations towards small energy scales (no more a power law below some lower threshold), a possible evolution of the lower cut-off $E_m$ with $\Delta$, or to the simplifying assumption of a brutal cut-off above $E^*$ in our calculations.

VII.    Divergence of an energy cut-off vs change in the power law exponent

Within our critical failure framework, we interpreted the observed evolution of the AE energy distributions towards failure (Fig. 3a of the main text) as truncated power law distributions with a constant exponent, but a diverging upper cut-off. This was fully consistent with the difference of exponent value between the distributions at failure, and the stress-integrated distributions (Fig. 3b of main text).

However, some authors reported, for the compressive failure of porous sandstone [63], or a mining collapse [64], a decreasing exponent towards failure. A brief look at the data of Fig. 3a (main text) might suggest at first glance a similar interpretation. To test this possibility, we performed additional statistical analyses as follows: We estimated, for each individual distribution (1 sample, 1 sensor), the exponent $\beta_E$, the associated uncertainty $\delta(\beta_E)$, and the p-value for the Kolmogorov-Smirnov test, following the maximum likelihood methodology proposed by [40]. Although there is a large scatter in the values of $\delta(\beta_E)$ and of the p-value from one distribution to another, we found, *systematically*, a positive correlation (0.15≤$R$≤0.45) between log($\Delta$) and $\delta(\beta_E)$ (meaning that the uncertainty on the exponent value increases as one goes away from failure), and a negative correlation (-0.35≤$R$≤-0.10) between the p-value and log($\Delta$) (meaning that the "pure" power law model is a worse and worse statistical model as one goes away from the failure). Taken altogether, this shows that our model is, in statistical terms only, better than the pure power law model to explain our data (not speaking about the underlying theoretical framework described in the main text). It has already been stressed that a misinterpretation of such data would suggest a (apparent) decreasing exponent as approaching failure [65].